\RequirePackage{rotating} 
\documentclass[]{jfm}
\graphicspath{{figs/}}

\usepackage{graphicx}
\usepackage{newtxtext}
\usepackage{newtxmath}
\usepackage{natbib}

\usepackage{hyperref}
\usepackage{amsmath}
\usepackage{tikz}
\usetikzlibrary{plotmarks}
\newcommand\marksymbol[2]{\tikz[#2,scale=1.2]\pgfuseplotmark{#1};} 
\usepackage{afterpage} 
\usepackage{rotating}
\hypersetup{
    colorlinks = true,
    urlcolor   = blue,
    citecolor  = black,
}
\newcommand{\RomanNumeralCaps}[1]
\linenumbers

\newcommand{\ub}{U_b}
\newcommand{\uw}{U_w}
\newcommand{\ul}{U^L}
\newcommand{\ud}{U^\Delta}
\newcommand{\ut}{u_\tau}
\newcommand{\ualp}{U_\alpha}

\newcommand{\uq}{U_\text{q}}

\newcommand{\aver}[1]{\left\langle #1 \right\rangle}

\newcommand{\pard}[2]{\frac{\partial #1}{\partial #2}}

\newcommand{\re}{\mathit{\Rey}}
\newcommand{\repi}{\mathit{\Rey}_\pi}
\newcommand{\ret}{\mathit{\Rey}_\tau}

\newcommand{\req}{\mathit{\Rey}_\text{q}}
\newcommand{\rew}{\mathit{\Rey}_w}
\newcommand{\reb}{\mathit{\Rey}_b}
\newcommand{\realp}{\mathit{\Rey}_\alpha}

\newcommand{\tw}{\tau_w}
\newcommand{\twt}{\tau_{w,t}}
\newcommand{\twb}{\tau_{w,b}}
\newcommand{\tws}{\tau_s}
\newcommand{\twa}{\tau_a}

\newcommand{\pit}{\Pi_t}
\newcommand{\piw}{\Pi_w}
\newcommand{\pip}{\Pi_p}
\newcommand{\phil}{\Phi^L}
\newcommand{\phid}{\Phi^\Delta}
\newcommand{\pp}{\mathcal{P}}
\newcommand{\ppl}{\mathcal{P}^L}
\newcommand{\ppd}{\mathcal{P}^\Delta}

\newcommand{\alp}{\alpha_P}
\newcommand{\alc}{\alpha_C}

\DeclareMathSymbol{\shortminus}{\mathbin}{AMSa}{"39}

\begin{document}

\title[Energy budgets in Couette and Poiseuille flows]{Global energy budgets \\ in turbulent Couette and Poiseuille flows}

\author[A. Andreolli, M. Quadrio \& D. Gatti]{
A\ls N\ls D\ls R\ls E\ls A\ls \ns
A\ls N\ls D\ls R\ls E\ls O\ls L\ls L\ls I\aff{1,2}, \ns 
M\ls A\ls U\ls R\ls I\ls Z\ls I\ls O\ls \ns 
Q\ls U\ls A\ls D\ls R\ls I\ls O\aff{2}
\and \ns
D\ls A\ls V\ls I\ls D\ls E\ls \ns 
G\ls A\ls T\ls T\ls I\aff{1}, \ns 
}
\affiliation{
\aff{1} Institute of Fluid Mechanics, Karlsruhe Institute of Technology, Kaiserstr. 10, 76131 Karlsruhe, Germany \\
\aff{2} Department of Aerospace Sciences and Technologies, Politecnico di Milano, via La Masa 34, 20156 Milano, Italy \\
[\affilskip]
}


\maketitle

\begin{abstract}
Turbulent plane Poiseuille and Couette flows share the same geometry, but produce their flow rate owing to different external drivers, pressure gradient and shear respectively. By looking at integral energy fluxes, we pose and answer the question of which flow performs better at creating flow rate. We define a flow {\em efficiency}, that quantifies the fraction of power used to produce flow rate instead of being wasted as a turbulent overhead; {\em effectiveness}, instead, describes the amount of flow rate produced by a given power. The work by Gatti \emph{et al.} (\emph{J. Fluid Mech.} vol.857, 2018, pp. 345--373), where the constant power input (CPI) concept was developed to compare turbulent Poiseuille flows with drag reduction, is here extended to compare different flows.
By decomposing the mean velocity field into a laminar contribution and a deviation, analytical expressions are derived which are the energy-flux equivalents of the FIK identity. These concepts are applied to literature data supplemented by a new set of direct numerical simulations, to find that Couette flows are less efficient but more effective than Poiseuille ones. The reason is traced to the more effective laminar component of Couette flows, which compensates for their higher turbulent activity. It is also observed that, when the fluctuating fields of the two flows are fed with the same total power fraction, Couette flows dissipate a smaller percentage of it via turbulent dissipation. A decomposition of the fluctuating field into large and small scales explains this feature: Couette flows develop stronger large-scale structures, which alter the mean flow while contributing less significantly to dissipation.
\end{abstract}

\label{sec:sym}


\maketitle

\section{Introduction}
\label{sec::introduction}

This work describes at an integral level the process by which turbulent Poiseuille and Couette flows --- which share the simple geometrical setting but possess intrinsic differences --- use a fraction of the external driving power to produce a flow rate, and dissipate the remainder via turbulence. The interest in these two prototypical flows resides in the fact that Poiseuille flows are pressure-driven, while Couette ones are powered by shear forces lumped at the wall.  

Both flows are wall-bounded, hence the relevance of viscous scaling: viscous or `plus' units are built with the wall-based velocity scale $\ut^* = \sqrt{\tw^*/\rho^*}$ (the asterisk denotes dimensional quantities, $\rho^*$ is the fluid density, and $\tw^*$ the wall-shear stress) and the kinematic viscosity $\nu^*$ of the fluid. However, viscous units sometimes fail at recovering universality when comparing different plane wall-bounded flows. For example, it is known that turbulence develops faster with $h^+$ in Couette flows, where $h^+ = h^* \ut^*/ \nu^*$ is the friction Reynolds number built with the friction velocity and the channel half height $h^*$. Indeed, the shear and wall-normal Reynolds stresses are known to saturate faster in Couette flows, and turbulence is sustained at lower $h^+$ \citep{orlandi-bernardini-pirozzoli-2015}; large near-wall structures have been observed in Couette flows at values of $h^+$ as low as $93$, while spectral peaks at low wavenumbers are encountered in Poiseuille flows at much higher $h^+$ (roughly, $h^+ = 5000$) and mostly in the central region of the channel \citep{lee-moser-2018}. 

As a consequence, one cannot simply resort to scaling when comparing different flows over a plane wall; for the comparison to be meaningful, prescribing the value of the Reynolds number is also necessary. It is common practice \citep[see, for instance,][]{monty-etal-2009,sillero-jimenez-moser-2013} to compare different wall-bounded flows at the same friction Reynolds number. However, not only this choice is discretionary, but also the definition itself of the friction Reynolds number contains arbitrariness in the choice of length scale \citep{jimenez-etal-2010}. The choice of $h^*$ made above seems reasonable, at least for Poiseuille flows \citep{jimenez-hoyas-2008}. The same convention is often adopted in literature for Couette flows as well, even though many argue that the channel full-width $2h^*$ would be better suited than $h^*$ as a length scale \citep{barkley-tuckerman-2007,lee-moser-2018}. 

To set up a proper comparison, the framework introduced by \cite{gatti-etal-2018} to describe global energy fluxes will be here extended. The framework was originally conceived to compare the same Poiseuille flow under different flow control strategies; here it will be generalized to compare flows that differ altogether. A criterion to set up a sensible comparison is needed: as in flow control, where one can compare at the same pressure gradient, the same flow rate or the same power input \citep{quadrio-frohnapfel-hasegawa-2016}, multiple possibilities exist, none of which can be excluded {\em a priori}. In the present context, the Constant Power Input (CPI) criterion will be shown to provide some advantages.



After introducing the flows of interest and the adopted notation in \S\ref{sec::notation}, the framework used to study the global energy budgets is presented in \S\ref{sec::geb}. It is then applied to a database of existing Direct Numerical Simulations, complemented by some newly produced ones, in \S\ref{sec::results}, where trends of energy budgets with different Reynolds numbers are considered. In \S\ref{sec::const_realp} a scale analysis addresses the contribution of the large Couette structures to the turbulent dissipation, and \S\ref{sec::conclusions} contains a concluding discussion.

\subsection{Notation and problem statement}
\label{sec::notation}

In this paper, an asterisk superscript $\cdot^*$ denotes dimensional quantities; non-dimensional ones will be written as bare symbols, except those scaled in viscous units, for which the conventional `plus' notation $\cdot^+$ will be employed. Let $\mathbf{u}^*$ be the velocity vector and $u^*,v^*,w^*$ its Cartesian components; by indicating with $\aver{\cdot}$ the temporal average, the usual Reynolds decomposition of the velocity field in its mean and fluctuating components is:
\begin{equation}
    \mathbf{u}^* = \mathbf{U}^* + \mathbf{u'}^*
\end{equation}
where $\mathbf{U}^* \equiv \aver{\mathbf{u}^*}$ is the mean velocity, with components $U^*$, $V^*$, $W^*$; the fluctuating velocity field $\mathbf{u'}^*$ (with its components ${u'}^*$, ${v'}^*$ and ${w'}^*$) is consequently defined. 

\begin{figure}
\centering
\includegraphics{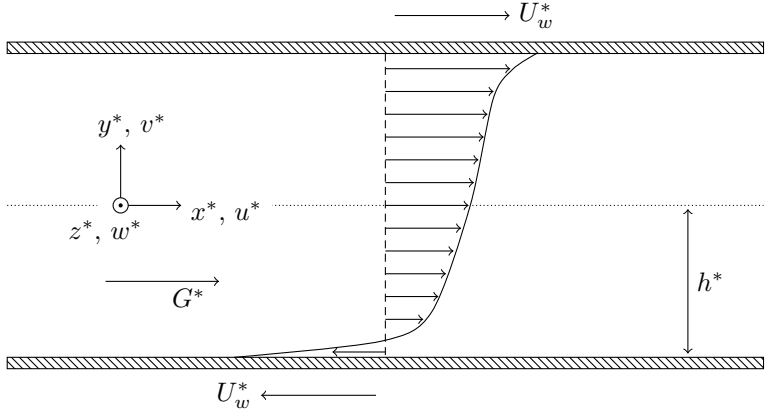}
\caption{Sketch of the flow and reference system.}
\label{fig::sketch}
\end{figure}

Let us now consider (see figure \ref{fig::sketch}) the statistically steady flow between two indefinite, parallel plates, forced by a pressure gradient and/or a relative movement of the plates. Let $h^*$ be half the gap between the plates; a system of Cartesian axes is located with origin at the mid-plane, so that the $y$ axis points in the wall-normal direction. The mean pressure gradient $\partial \aver{P^*}/\,\partial x^* = -G^*$ drives the flow in the streamwise ($x$) direction; without loss of generality, we assume $G^*>0$. The two walls move in the streamwise direction with velocity $\pm U_w^*$. This is the combined Couette--Poiseuille flow, which reduces to the simple Couette flow for $G^*=0$, and to the simple Poiseuille flow for $\uw^* = 0$. The bulk velocity is written as $U_b^*$; although for the simple Couette flow $U_b^*=0$, the wall velocity $U_w^*$ can be regarded as the bulk velocity once the flow is observed in a reference frame where one of the two walls is at rest. Hence, the flow rates realized by simple Poiseuille and Couette flows are $U_b^*$ and $U_w^*$ respectively.

A (yet unspecified) non-dimensionalisation employing $h^*$ as the length scale defines a Reynolds number $\re$. The balance of the kinetic energy $U^2 = 0.5 U_i U_i$ of the mean flow (MKE) can thus be written in a dimensionless form specialized for the plane channel as \citep{pope-2000}:
\begin{align}
\label{eq::local_mke}
    \frac{\Bar{D}}{Dt}\frac{U^2}{2}\;\;\; = 
    \underbrace{\;\;\; \vphantom{\frac{d\aver{P}}{dy}}G U\;\;\;}_{\text{pumping power}}
    + \underbrace{\;\;\;-r(y) \frac{dU}{dy}\;\;\;}_{\text{turb. production}}
    + \underbrace{\;\;\;\frac{d}{dy}\left(r(y)U\right)\;\;\;}_{\text{turb. transport}} \notag \\
    + \underbrace{\;\;\;\frac{1}{\re} \frac{d}{dy} \left( U \frac{dU}{dy} \right)\;\;\;}_{\text{viscous diffusion}}
    - \underbrace{\;\;\;\frac{1}{\re} \left(\frac{dU}{dy}\right)^2\;\;\;}_{\text{dissipation}}
\end{align}
where $r(y) = -\aver{u'v'}$ is the wall-normal profile of the Reynolds shear stress.
In this balance, the pumping power acts as a source term, while production of turbulent kinetic energy and dissipation both act as a sink. Two (turbulent and viscous) transport terms are also present. The balance of turbulent kinetic energy $k=0.5\aver{u_i' u_i'}$ (TKE) reads:
\begin{align}
\label{eq::local_tke}
    \frac{\Bar{D}k}{Dt} \;\;\; = 
    \underbrace{\;\;\; r(y)\frac{dU}{dy} \;\;\;}_{\text{turb. production}}
    + \underbrace{\;\;\;- \frac{1}{2}\frac{d}{dy} \aver{v'u_i' u_i'}\;\;\;}_{\text{turb. transport}}
    +\underbrace{\;\;\;- \frac{d}{dy} \aver{P'v'}\;\;\;}_{\text{pressure transport}} \notag\\
    + \underbrace{\;\;\; \frac{1}{\re} \frac{d^2k}{dy^2} \;\;\;}_{\text{viscous diffusion}}
    - \underbrace{\;\;\;\frac{1}{\re} \aver{\pard{u_i'}{x_k}\pard{u_i'}{x_k}}\;\;\;}_{\text{turb. dissipation}}
\end{align}
where this time production acts as a source, and three (turbulent, pressure and viscous) transport terms are present together with a sink (dissipation). Notice that in both cases the pseudo-dissipation formulation (also referred to as isotropic dissipation) has been used instead of the thermodynamically correct one \citep{bradshaw-perot-1993}.

Equations \eqref{eq::local_mke} and \eqref{eq::local_tke} are integrated in the wall-normal direction over the full channel height and then halved to obtain balances per unit wet surface for MKE and TKE. Notice that all the divergence terms (except for the mean viscous diffusion) integrate to zero. This results in:
\begin{align}
\label{eq::old_mke}
	& & &\text{MKE:} & \pit - \Phi - \pp &= 0 & \\
\label{eq::old_tke}
    & & &\text{TKE:} & \pp - \epsilon &= 0 & 
\end{align}
where $\pit$ is the total power input to the flow, and contains two contributions, $\pip$ and $\piw$. The former represents the power provided by the pressure gradient and arises from the integration of the pumping power term; the latter is the power of the external forces applied to the walls to keep them at a speed $\uw$, and arises from integration of the mean viscous diffusion term. By indicating the mean shear stress $(dU/dy)/\re$ at the top and bottom walls as $\twt$ and $\twb$ respectively, it can be further shown that $G = \left( \twb - \twt \right) / 2$. Hence:
\begin{alignat}{3}
	\label{eq::pit}
	&\pit &&= \pip + \piw & \\
	\label{eq::pip}
	&\pip &&= \frac{1}{2} \left( \twb - \twt \right) \ub &&= \twa \ub \\
	\label{eq::piw}
	&\piw &&= \frac{1}{2} \left( \twt + \twb \right)\uw &&= \tws \uw \, ,
\end{alignat}
where $\tws$ and $\twa$ are the symmetric and antisymmetric wall shear stress respectively. Both velocity scales appearing in the definition of power ($\ub$ and $\uw$ in equations \eqref{eq::pip} and \eqref{eq::piw}) will be referred to as \emph{flow rates}, since, as explained above, they represent flow rate in simple Poiseuille and Couette flows.

Eq. \eqref{eq::old_mke} shows that part of the power $\pit$ is wasted by the dissipation $\Phi$ of the mean flow, given by integration of the corresponding term:
\begin{equation}
	\Phi = \frac12 \int_{-1}^1 \frac{1}{\re} \left(\frac{dU}{dy}\right)^2\,dy\\
\end{equation}
and the remainder is transformed into turbulent kinetic energy by turbulent production:
\begin{equation}
	\mathcal{P} = \frac12 \int_{-1}^1 r(y) \frac{d U}{dy} \,dy
\end{equation}
Finally, turbulent dissipation $\epsilon$ (arising from the integration of the corresponding term) degrades the energy fed from the mean flow to turbulent fluctuations:
\begin{equation}
\epsilon = \frac12 \int_{-1}^1 \frac{1}{\re} \aver{\pard{u_i'}{x_k}\pard{u_i'}{x_k}} \,dy
\end{equation}

\section{The Constant-Power-Input framework for the Couette--Poiseuille flow family}
\label{sec::geb}

A conceptual framework was designed by \cite{gatti-etal-2018} to rationally compare Poiseuille flows with and without flow control in terms of integral energy fluxes under constant power input (CPI). That analysis will be here extended to the Couette configuration, to enable a CPI comparison between different flows. The general combined Couette--Poiseuille flow family is discussed first, and results for simple Couette and simple Poiseuille flows will be recovered later.

The starting point is choosing the velocity scale for a meaningful non-dimensionalisation, once the length scale $h^*$ has been set. We opt for the so-called power velocity $U_\pi^*$:
\begin{equation}
U_\pi^* = \left(\frac{\pit^*}{\rho^*}\right)^{1/3}
\label{eq:power-velocity}
\end{equation}
which in its definition resembles the friction velocity, except that the total power input per unit wetted area is used instead of the wall shear. This naturally leads to the definition of a power-based Reynolds number $\repi = h^* U_\pi^* / \nu^*$. Using this non-dimensionalisation is equivalent to expressing all energy fluxes as fractions of $\pit^*$; obviously, $\pit=1$. Notice that the definition (\ref{eq:power-velocity}) of the power velocity is more general than the one $U_{\pi, Pois}^*$ given in previous works \citep{hasegawa-quadrio-frohnapfel-2014,gatti-etal-2018} and valid for Poiseuille flows only; the following relation holds:
\begin{equation}
	U_\pi^* = \left( 3 \frac{\nu^*}{h^*} \right)^{1/3} \left( U_{\pi, Pois}^* \right)^{2/3} .
\end{equation}
Under the general definition (\ref{eq:power-velocity}), any two Couette--Poiseuille flows with the same values of $\repi$ are driven by the same power input. The definition is also independent from the reference frame, and immediately applies to unusual flow configurations such as Poiseuille flows with no flow rate \citep{tuckerman-etal-2014} or experimental results for Couette and Couette--Poiseuille flows \citep{kawata-alfredsson-2019, klotz-etal-2021}. 

Thanks to the definitions \eqref{eq::pip}, \eqref{eq::piw} and \eqref{eq:power-velocity}, the total power \eqref{eq::pit} can be recast in terms of Reynolds numbers:
\begin{equation}
	\repi^3 = (h^+_s)^2\rew + (h^+_a)^2\reb \, ,
\label{eq::repitauk_mixed}
\end{equation}
where $h^+_s = h^* \sqrt{\tws^*/\rho^*}/\nu^*$ and $h^+_a = h^*\sqrt{\twa^*/\rho^*}/\nu^*$ are the friction Reynolds numbers defined with the friction velocity descending from the symmetric and anti-symmetric wall shear stresses, while $\reb = h^* \ub^*/\nu^*$ and $\rew = h^*\uw^*/\nu^*$ are Reynolds numbers where the velocity scale is given by the flow rates.

In addition to the power Reynolds number, a second parameter is needed for the characterisation of a Couette--Poiseuille flow. The two usually employed parameters \citep{eltelbany-reynolds-1980,eltelbany-reynolds-1981, nakabayashi-etal-2004} are the friction Reynolds number $h^+_b$ of the bottom wall and the flow parameter $\theta$:
\begin{equation}
	h^+_b = \frac{h^* \sqrt{\twb^*/\rho^*}}{\nu^*}\text{,} \;\;\;\;\;\;\;\;\;	\theta = \frac{h^*}{\twb^*} \frac{d \tau^*}{dy^*} = - \frac{\twa^*}{\twb^*}
\end{equation}
where $\tau^*$ is the total shear stress. In the present, power-focused approach, we use instead $\repi$ and the pumping power share $a = \pip / \pit$. Hence, a pair $(\repi, a)$ or equivalently $(h_b^+,\theta)$ fully describes the state of a Couette--Poiseuille flow. Thanks to equations \eqref{eq::pit}, \eqref{eq::pip}, \eqref{eq::piw} and \eqref{eq::repitauk_mixed}, the two parameter sets can be related as:
\begin{equation}
	\begin{cases}
		\;\repi = \left( \, (1+\theta)(h^+_b)^2\rew -\, \theta(h^+_b)^2\reb \, \right)^{1/3}\\
		\;a = \left( 1 - \frac{1+\theta}{\theta} \frac{\rew}{\reb} \right)^{-1}
	\end{cases}
\end{equation}
where $\rew$ and $\reb$ implicitly depend on $h^+_b$ and $\theta$ (an explicit relation cannot be obtained due to the closure problem of turbulence).

\subsection{The extended Reynolds decomposition}

The next step of our analysis consists in extending the classic Reynolds decomposition, as in \cite{gatti-etal-2018}. After splitting the velocity field into a mean component $U(y)$ and a fluctuating part, the former is further decomposed as the sum of a Stokes (laminar) and a deviation part:
\begin{equation}
U(y) = \ul(y) + \ud(y)
\label{eq::ext_re_dec}
\end{equation}
where $\ul$ is the Stokes solution of the problem under consideration that achieves the same flow rate as the turbulent one, i.e 
\begin{equation}
	\ul(y) =  \uw\, y + \frac{3}{2}\,\ub\left( 1 - y^2 \right) .
\end{equation}
Hence, the deviation profile $\ud$ is zero at the wall and has zero integral; in other words, it does not contribute to either $\ub$ or $\uw$. For a generic parallel flow with constant cross-section, $\ul$ is the solution that minimizes the power required to generate a given flow rate \citep{bewley-2009, fukagata-sugiyama-kasagi-2009}. 

\begin{figure}
\centering
\includegraphics[]{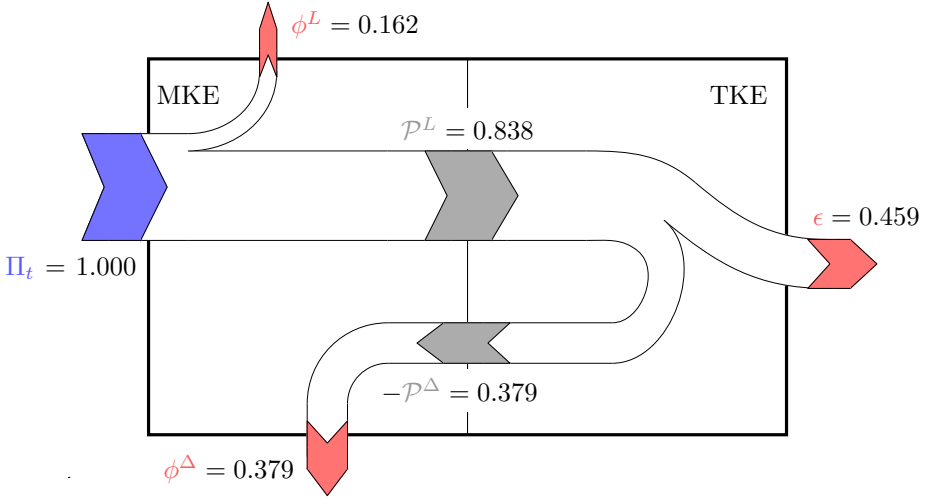} 
\caption{Extended energy box for a Couette flow at $h^+ = 102$, with numerical values of the integral terms expressed in power units.}
\label{fig::ebox}
\end{figure}

The wall-shear stresses can also be decomposed into the sum of a laminar and a deviation part, i.e. $\tws = \tws^L + \tws^\Delta$ and $\twa = \twa^L + \twa^\Delta$. As already observed in \cite{gatti-etal-2018} for Poiseuille flows, the extended Reynolds decomposition also decouples the integral power budget terms:  
\begin{align}
	\label{eq::def_phil_phid}
	\Phi \;&=\; \phil + \phid \;=\; \frac12 \int_{-1}^1 \frac{1}{\repi} \left(\frac{d\ul}{dy}\right)^2\,dy \;+\; \frac12 \int_{-1}^1 \frac{1}{\repi} \left(\frac{d\ud}{dy}\right)^2\,dy\\
	\pp \;&=\; \ppl + \ppd \;=\; \frac12 \int_{-1}^1 r(y) \frac{d\ul}{dy} \,dy \;+\; \frac12 \int_{-1}^1 r(y) \frac{d\ud}{dy} \,dy
\end{align}
Moreover, it can be proved that $\ppd = -\phid <0$: hence the positive production term $\ppl$ transfers energy from the mean to the fluctuation field, whereas the deviation production $\ppd$ acts in the opposite direction to be a sink for TKE. Eventually, thanks to the extended decomposition, equations \eqref{eq::old_mke} and \eqref{eq::old_tke} can be rewritten as:
\begin{align}
	\label{eq::mke_bal}
	&&&\text{MKE:} & 1 - \phil - \ppl &= 0 &\\
	\label{eq::tke_bal}
	&&&\text{TKE:} & \ppl - \phid - \epsilon &= 0 &
\end{align}
where the total power input term $\pit$ becomes unity as a result of the power-based scaling. The interpretation of the various integral terms is now straightforward: of the power input $\pit$, the dissipated fraction $\phil$ is the smallest amount of power required to achieve the given flow rate for the flow under consideration \citep{gatti-etal-2018}. The remaining dissipation is caused by the presence of turbulence, and exits the system either as turbulent ($\epsilon$) or deviational ($\phid$) dissipation; the two contributions sum up to $\ppl$. The whole process is conveniently represented by the energy box \citep{quadrio-2011,ricco-etal-2012}, drawn in figure \ref{fig::ebox} for a Couette flow at $h^+=102$ (one of the cases discussed below). The various components of the balance are a function of the flow type and Reynolds number.

\subsection{Analytical expressions for the energy fluxes}
\label{sec::cpi_expressions}

Most of the quantities defined above can be expressed analytically in terms of the Reynolds number $\repi$, the pumping power share $a=\Pi_p/\Pi_t$, and the following two weighted integrals of the Reynolds shear stress $r(y)$:
\begin{equation}
\label{eq::nondim_integrals}
\alp  = \frac12 \int_{-1}^1 - y r(y) dy , \qquad 
\alc  = \frac12 \int_{-1}^1     r(y) dy .
\end{equation}
Due to the anti-symmetric weight in a symmetric integration domain, $\alp$ is only determined by the anti-symmetric part of the Reynolds stress, and is therefore zero for Couette flows; conversely, $\alc$ depends on the symmetric part alone, and is zero for Poiseuille ones. The weight functions are the wall-normal derivative of the laminar profile $U^L(y)$, normalized by their wall value.

Integration of the momentum equation written for the deviation component (not shown for brevity) leads to $\tws^\Delta = \alc$ and $\twa^\Delta = 3 \alp$: once again, the symmetric and antisymmetric parts of $r(y)$ separate. Although separable, they are mutually dependent due to the non-linear nature of the Navier--Stokes equations. Separation is also possible for the flow rates, with $\uw$ and $\ub$ associating to $\alc$ and $\alp$ respectively:
\begin{equation}
  \uw = \frac{\repi \alc}{2} \left( \sqrt{1 + \frac{4(1-a)}{\repi\alc^2}} - 1 \right) , \qquad
  \ub = \frac{\repi \alp}{2} \left( \sqrt{1 + \frac{4a}{3\repi\alp^2}} - 1 \right) .
\end{equation}
These expressions are the CPI equivalent of the FIK identity originally derived by \cite{fukagata-iwamoto-kasagi-2002}. 
Furthermore, an expression for $\ppl$ is obtained:
\begin{equation}
	\ppl = \frac{\repi\alc^2}{2} \left( \sqrt{1 + \frac{4(1-a)}{\repi\alc^2}} - 1 \right) + \frac{3\repi \alp^2}{2}\left( \sqrt{1 + \frac{4a}{3 \repi\alp^2}} -1 \right)
	\label{eq::pl_combined}
\end{equation}
and, since from equation \eqref{eq::mke_bal} $\phil = 1 - \ppl$, $\phil$ becomes analytically known too. Lastly, an expression for $\phid$ is needed; first, the following integral is defined: 
\begin{equation}
\beta = \frac12 \int_{-1}^1 r^2(y) dy .
\end{equation}
Then, $\phid$ can be written as:
\begin{equation}
\phid = \repi \left( \beta - \alc^2 - 3 \alp^2 \right) .
\label{eq::expr_phid}
\end{equation}
The Cauchy--Schwarz inequality proves that this quantity is always non-negative.

\subsection{Simple Couette and Poiseuille flows}
\label{sec::isolated}

Expressions for simple Couette and Poiseuille flows are obtained by setting $a = 0$ or $a = 1$ respectively. Simple Couette flows have an antisymmetric mean velocity profile, hence shear and Reynolds stresses are symmetric: $\tw = \tws$ and $\twa = 0$, with $\alp=0$ and $\ub=0$. Conversely, simple Poiseuille flows have a symmetric mean velocity profile, hence shear and Reynold stresses are antisymmetric: $|\tw| = \twa$ and $\tws = 0$, with $\alc=0$ and $U_w=0$. For both, equation \eqref{eq::repitauk_mixed} can be simplified to:
\begin{equation}
	\repi^3 = \left(h^+\right)^2\req
	\label{eq::repitauk}
\end{equation}
where $\req$ is the flow-rate-based Reynolds number, i.e. $\req = \rew$ for Couette flows and $\req = \reb$ for Poiseuille ones.


Equation \eqref{eq::pl_combined} can be further simplified into an expression for $\ppl$ (or $\phil$) that is valid for both simple Couette and Poiseuille flows. To this purpose, a new Reynolds number is introduced, based on a new velocity scale $\ualp^*$. The latter is given by the ratio between the weighted integral $\alpha^*$ of the Reynolds shear stress and the flow rate $\uq^*$:
\begin{equation}
	\ualp^* = \frac{\alpha^*}{\uq^*} \text{,} \qquad \alpha^* = \frac{1}{2h^*} \int_{-h^*}^{h^*} r^*(y^*) \psi(y^*)  \,dy^* \text{,}
\end{equation}
where $\psi(y^*) = \left(\mathrm{d}\ul /  \mathrm{d}y \right) / \left(\mathrm{d}\ul /  \mathrm{d}y \right)_{y=0}$ is the same non-dimensional weight of equation \eqref{eq::nondim_integrals}, such that $\alpha^* = \alc^*$ for Couette flows, or $\alpha^* = \alp^*$ for Poiseuille ones. The new Reynolds number $\realp = h^* \ualp^* / \nu^*$ is consequently defined; $\realp$ already appears in the FIK identity for Couette flows \citep{kawata-alfredsson-2019}, which can be cast as
\begin{equation}
	\realp = \frac{\ret^2}{\rew} - 1 \text{,}
\end{equation}
and the one for $\ub^+$ of Poiseuille flows \citep{marusic-joseph-mahesh-2007}, 
\begin{equation}
	\realp = \frac{\ret^2}{3 \reb} - 1 \text{.}
\end{equation}
Eventually, the aforementioned fluxes $\ppl$ and $\phil$ can be written as functions of $\realp$:
\begin{equation}
\ppl = \frac{1}{1 + 1/\realp}, \qquad \phil = \frac{1}{1 + \realp} .
\label{eq::ppl_realp}
\end{equation}

The laminar production $\ppl$ expresses the fraction of external power wasted because of turbulence, and is of particular interest as it quantifies the overhead in producing flow rate from a given power. Due to the used non-dimensionalisation, the dissipation $\phil$ of the laminar component is the ratio between the power $\Pi_L^*$ required by the Stokes solution and the total power input $\pit^*$,
\begin{equation}
	\phil = \frac{\Pi_L^*}{\pit^*}\text{,}
	\label{eq::def_efficiency}
\end{equation}
with $\Pi_L$ being the theoretical minimum power needed to achieve a given flow rate. Therefore, $\phil$ represents an efficiency: the closer $\phil$ to one, the closer the flow to the ideal situation where the whole power is spent to produce flow rate. Summing up, the whole MKE box is determined by the value of $\realp$; it remains to be determined when a given $\realp$ is obtained, and what happens to the TKE box. This requires additional information from DNS datasets because of the unknown distribution $r(y)$, and will be addressed below.

\section{Numerical results}
\label{sec::results}

The discussion that follows is based on a set of DNS simulations of turbulent Poiseuille and Couette flows. The dataset includes simulations carried out for the present work as well as published data from \cite{lee-moser-2015, orlandi-bernardini-pirozzoli-2015, gatti-quadrio-2016, gatti-etal-2018, lee-moser-2018}. The new set of simulations, carried out with the code described in \cite{luchini-quadrio-2006}, includes two Couette flows with $h^+ \simeq 100$ and $500$, as well as Poiseuille flows at $h^+ = 100$, $150$, $316$ and $500$, selected to provide additional data points when needed; details of the various cases and their spatial discretization are reported in table \ref{tab::DNS_inventory}. As for the Couette simulations, a streamwise domain length long enough to accommodate large-scale motions in the core of the channel has been used; following \cite{orlandi-bernardini-pirozzoli-2015}, a value of $12\pi h^*$ is chosen here for the low Reynolds number simulation, while $16\pi h^*$ is used for the higher Reynolds number. These domain lengths have limited effects on one-point statistics \citep{lee-moser-2018}, mainly affecting two-points statistics and the spatial orientation of structures \citep{komminaho-etal-1996}, neither of which is of primary interest in this study.

  \begin{sidewaystable}
  \centering
  \rule{\textwidth}{0.4pt}
  \phantom{\rule{\textwidth}{0.3cm}} 
  \begin{tabular}{c c c c c c c c c c c c c}
  \hspace{.3cm} & Dataset & Flow & $L_x^*/(\pi h^*)$ & $L_z^*/(\pi h^*)$ & $\Delta x^+$ & $\Delta z^+$ & $\Delta y_w^+$ & $\Delta y_{c}^+$ & $h^+$ & $\repi$ & $\req$ & $\realp$ \\ 
  \hline
  \marksymbol{triangle*}{red}  & \cite{orlandi-bernardini-pirozzoli-2015} & C & 18 & 8 & 7.6 & 4.8 & 0.03 & N.A. & 171 & 445 & 3000 & 8.556 \\          
  \marksymbol{triangle*}{red}  & \cite{orlandi-bernardini-pirozzoli-2015} & C & 18 & 8 & 7.2 & 5.1 & 0.04 & N.A. & 260 & 688 & 4800 & 12.728 \\         
  \marksymbol{triangle*}{red}  & \cite{orlandi-bernardini-pirozzoli-2015} & C & 18 & 8 & 7.0 & 5.0 & 0.06 & N.A. & 508 & 1373 & 10134 & 24.296 \\       
  \marksymbol{triangle*}{red}  & \cite{orlandi-bernardini-pirozzoli-2015} & C & 18 & 8 & 6.8 & 4.8 & 0.08 & N.A. & 986 & 2744 & 21334 & 44.281 \\       
  \marksymbol{pentagon*}{red}  & \cite{lee-moser-2018}                    & C  & 100 & 5 & 9.5 & 5.1 & 0.03 & 2.4 & 93 & 235 & 1500 & 4.623 \\         
  \marksymbol{pentagon*}{red}  & \cite{lee-moser-2018}                    & C  & 100 & 5 & 11.2 & 4.5 & 0.03 & 3.7 & 219 & 578 & 3995 & 10.864 \\       
  \marksymbol{pentagon*}{red}  & \cite{lee-moser-2018}                    & C  & 100 & 5 & 10.3 & 5.1 & 0.04 & 6.3 & 501 & 1360 & 9995 & 23.817 \\     
  \marksymbol{|}{cyan, very thick}         & \cite{gatti-etal-2018}                   & P & 4 & 2 & 9.8 & 4.9 & 0.47 & 2.6 & 200 & 502 & 3177 & 3.190 \\           
  \marksymbol{oplus*}{cyan}    & \cite{gatti-quadrio-2016}                & P & 4 & 2 & 12.3 & 6.1 & 0.97 & 7.1 & 1001 & 2715 & 19993 & 15.573 \\      
  \marksymbol{diamond*}{cyan}  & \cite{lee-moser-2015}                    & P & 8 & 3 & 4.5 & 3.1 & 0.07 & 3.4 & 182 & 456 & 2829 & 2.872 \\           
  \marksymbol{diamond*}{cyan}  & \cite{lee-moser-2015}                    & P & 8 & 3 & 8.9 & 5.0 & 0.02 & 4.5 & 544 & 1435 & 9952 & 8.823 \\          
  \marksymbol{diamond*}{cyan}  & \cite{lee-moser-2015}                    & P & 8 & 3 & 10.9 & 4.6 & 0.02 & 6.2 & 1000 & 2715 & 19930 & 15.681 \\      
  \marksymbol{diamond*}{cyan}  & \cite{lee-moser-2015}                    & P & 8 & 3 & 12.2 & 6.1 & N.A. & N.A. & 1995 & 5572 & 43378 & 29.543 \\     
  \marksymbol{diamond*}{cyan}  & \cite{lee-moser-2015}                    & P & 8 & 3 & 12.7 & 6.4 & 0.50 & 10.3 & 5186 & 14980 & 124863 & 70.357 \\   
  \end{tabular}
  \vspace{.3cm}

  \begin{tabular}{c c c c c c c c c c c c c c}
  \hspace{.3cm} & Flow & $L_x^*/(\pi h^*)$ & $L_z^*/(\pi h^*)$ & $\Delta x^+$ & $\Delta z^+$ & $\Delta y_w^+$ & $\Delta y_{c}^+$ & $\Delta t u_\tau^* / h^*$ & $N_s$ & $h^+$ & $\repi$ & $\req$ & $\realp$ \\ 
  \hline
  \marksymbol{asterisk}{red, thick}        & C & 12 & 4 & 10.0 & 5.0 & 0.49 & 2.6 & 0.305 & 408 & 102 & 258 & 1667 & 5.215 \\      
  \marksymbol{asterisk}{red, thick}        & C & 16 & 8 & 12.5 & 6.2 & 0.99 & 7.2 & 0.5 & 223 & 507 & 1376 & 10133 & 24.40 \\
  \marksymbol{Mercedes star}{cyan, thick}  & P & 4  & 2 & 9.8  & 4.9 & 0.48 & 2.6 & 1.00  & 600 & 100 & 245 & 1467 & 1.270 \\      
  \marksymbol{Mercedes star}{cyan, thick}  & P & 4  & 2 & 7.4  & 3.7 & 0.72 & 3.9 & 1.81  & 71  & 150 & 373 & 2296 & 2.259 \\      
  \marksymbol{Mercedes star}{cyan, thick}  & P & 4  & 2 & 10.4 & 5.3 & 0.50 & 2.7 & 0.76  & 152 & 316 & 812 & 5370 & 5.191 \\      
  \marksymbol{Mercedes star}{cyan, thick}  & P & 4  & 2 & 12.3  & 6.1 & 0.97 & 7.1 & 1.00  & 186  & 500 & 1314 & 9085 & 8.177 \\    
  \end{tabular}
  \vspace{.3cm}
  \caption{DNS datasets for Poiseuille (P) and Couette (C) flows, including published (top) as well as new (bottom) simulations. $L_x^*$ and $L_z^*$ are the domain lengths in the streamwise and spanwise directions, with the corresponding $\Delta x^+$ and $\Delta z^+$ resolutions in wall units. $\Delta y_w^+$ is the wall-normal resolution at the wall in viscous units, and $\Delta y_c^+$ represents the same quantity at the centerline. Finally, $\Delta t$ and $N_s$ are the sampling time and the number of samples used for averaging. Additionally, the table introduces the color scheme and symbols used later.} 
  \label{tab::DNS_inventory}
  \rule{\textwidth}{0.4pt}
\end{sidewaystable}

\subsection{Comparison at constant $Re_\tau$}

\begin{figure}
\centering
\includegraphics{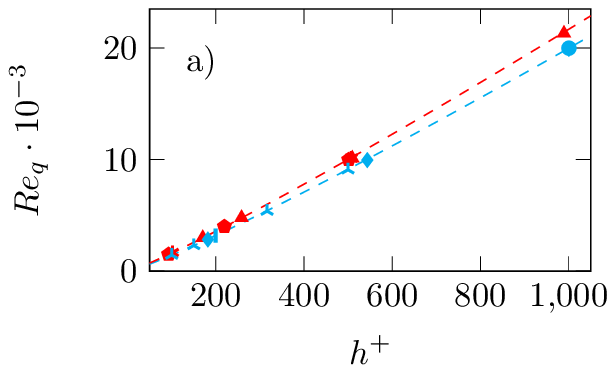}
\includegraphics{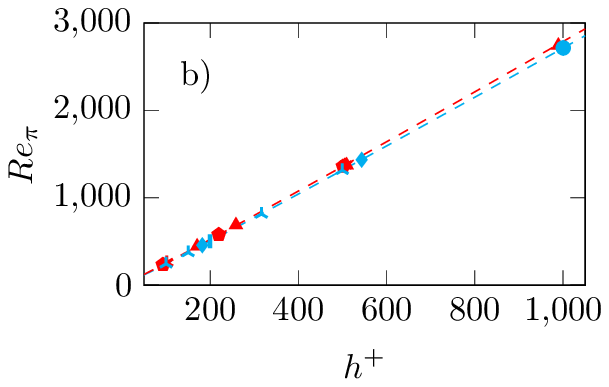}
\includegraphics{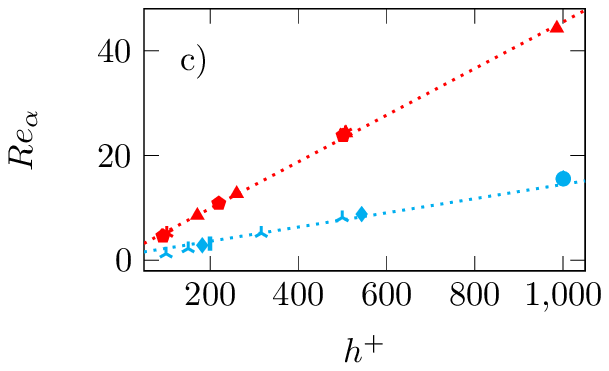}
\caption{Dependence of the Reynolds numbers $Re_q$, $Re_\pi$ and $Re_\alpha$ on the friction Reynolds number $h^+$, for Couette (red) and Poiseuille (blue) flows. The dashed lines in the panels above indicate analytical fits, whereas the dotted lines in the lower panel are an empirical linear fit. For colors and symbols, refer to table \ref{tab::DNS_inventory}.}
\label{fig::interpolations}
\end{figure}

First, the relationship among the various Reynolds numbers is discussed, and figure \ref{fig::interpolations} plots their variation with $h^+$. Panel (a) shows that $Re_q$ varies with $h^+$ quite similarly in the two flows; the similarity is even more striking when the variation of $\repi$ is considered, as shown in panel (b). In both cases, available data shows the expected nearly linear increase. For example, \cite{abe-antonia-2016} linked $h^+$ and $\reb$ for Poiseuille flows by assuming that deviations from the logarithmic law in the mean velocity profile near the mid-plane of the channel are negligible; the same approximation was found to be valid for Couette flows as well \citep{orlandi-bernardini-pirozzoli-2015}. The underlying functional form fitted to the present data yields:
\begin{alignat}{2}
&\frac{\reb}{h^+} = 2.85 + 2.48\ln(h^+) \label{eq::interp_poi} \\
&\frac{\rew}{h^+} = 5.17 + 2.39\ln(h^+) \label{eq::interp_cou}
\end{alignat}
and the use of \eqref{eq::repitauk} leads to analogous expressions for $\repi$ vs $h^+$. In both cases, the approximations are satisfactory. The most interesting result, however, is contained in panel (c) of figure \ref{fig::interpolations}, where the change of $\realp$ with $h^+$ is shown. Considerable quantitative differences are seen between Couette and Poiseuille flows: the slope of a linear fit to Couette data is approximately three times larger than the same slope for Poiseuille flows.

Comparing Couette and Poiseuille flows at the same $h^+$ is the most natural choice, used several times in the past (see \S\ref{sec::introduction}). This comparison yields quite a similar flow rate, and does not immediately reveal why a Couette flow possesses an increased level of turbulent activity with respect to a Poiseuille one at the same $h^+$. However, under such condition, $\realp$ differs considerably between the two flows,  suggesting a new and potentially informative comparison.

\subsection{Comparison at constant $\repi$}

\begin{figure}
\centering
\includegraphics{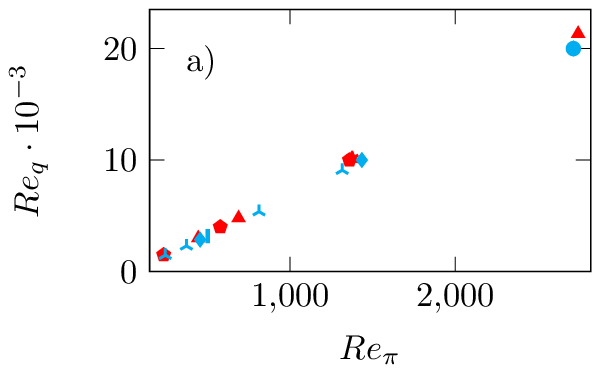}
\includegraphics{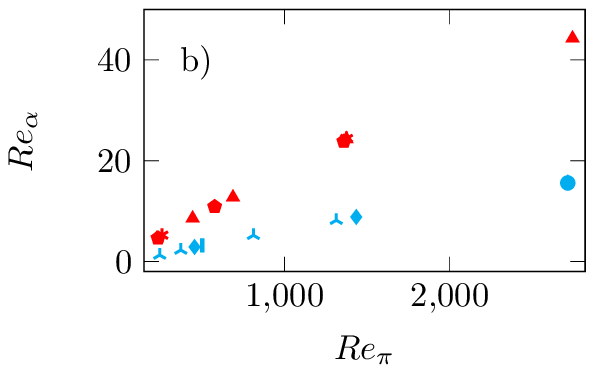}
\caption{Reynolds numbers $\req$ and $\realp$ against $\repi$. For colors and symbols, refer to table \ref{tab::DNS_inventory}.}
\label{fig::constant_power}
\end{figure}

\begin{figure}
\centering
\includegraphics{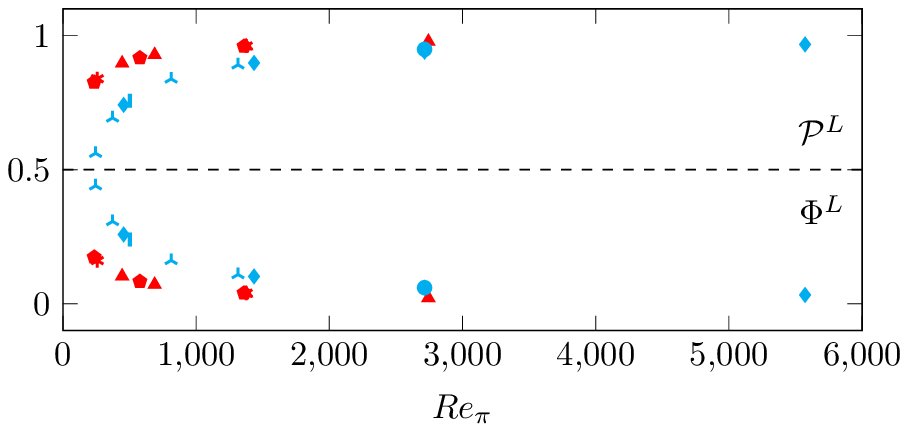}
\caption{Plot of $\ppl$ and $\phil=1-\ppl$ against $\repi$. For colors and symbols, refer to table \ref{tab::DNS_inventory}.}
\label{fig::actual_realp}
\end{figure}

Before addressing the comparison at the same $\realp$, which will provide a better indication of turbulent activity, we compare Couette and Poiseuille flows at constant power input. The relationship between $Re_\pi$ and $h^+$ has been already shown in figure \ref{fig::interpolations} (b) to be nearly linear for the available data, and to not depend on the flow type.  $\req$ and $\realp$ are plotted against $\repi$ in figure \ref{fig::constant_power} (a,b), respectively. The former panel allows to asses the effectiveness $\req(\repi)$ of the flow, i.e. the amount of flow rate $\req$ produced out of a given power input $\repi$. Data for Poiseuille and Couette flows almost collapse, meaning that they are similarly effective. The maringal difference between the two flows indicates that Couette flows are slightly more effective than Poiseuille ones on the whole range of available data. This was already pointed out by \cite{orlandi-bernardini-pirozzoli-2015}, and seems to be at odds with many evidences in literature of a higher turbulent activity in Couette flows, e.g. from the same authors (see \S\ref{sec::introduction} for more). Such higher turbulent activity is here confirmed in  panel (b): when the two flows are compared under the same power input, Couette flows achieve a much higher $\realp$ than Poiseuille ones. In view of the discussion in \S\ref{sec::isolated}, this implies that Couette flows exhibit a larger turbulent overhead $\ppl$, hence a lower efficiency $\phil$. Figure \ref{fig::actual_realp} better shows how efficiency $\phil$ and its complement overhead (or inefficiency) $\ppl$ change with $\repi$; it is clearly seen that the turbulent overhead is larger in Couette flows than in Poiseuille ones for a given power input.

Provided that a Couette flow is more turbulent than a Poiseuille one at CPI and that turbulence has an adverse effect, the reason for the better effectiveness of the Couette case has to be sought in its Stokes solution. The lower efficiency $\phil$ means that the Stokes component $\ul$ of a generic Couette flow is fed with a smaller fraction of the total power input; still, this $\ul$ requires less power than its Poiseuille counterpart to achieve a given flow rate, hence compensating for the lower power supply. In other words, the concept of efficiency -- after equation \eqref{eq::def_efficiency} -- addresses performance with respect to the ideal case of the flow under consideration, while effectiveness $\req(\repi)$ does that in absolute terms.

Another significant difference between the two flows is the presence of energetic, large-scale structures in Couette flows \citep{lee-moser-2018, pirozzoli-bernardini-orlandi-2014, kitoh-umeki-2008}. These essentially realize an inertial mechanism \citep{papavassiliou-hanratty-1997} that transfers momentum from one wall to the other -- or, in other words, produce flow rate. While they surely do not provide a better performance with respect to the optimal case of a Stokes solution, their comparison with smaller-scale turbulence is not trivial. Large-scale rolls in Couette flows can either be suppressed or energised by a Coriolis force; in the first case, drag reduction is observed at constant flow rate \citep{komminaho-etal-1996}, in the latter a drag increase is obtained instead \citep{kawata-alfredsson-2016b, bech-andersson-1996}, suggesting that large-scale structures have a negative impact on effectiveness. \cite{degiovannetti-hwang-choi-2016} also reported a degrading effect of large-scale structures on flow performance -- albeit in Poiseuille flows. In light of this, the Stokes solution can be considered the main responsible for the larger flow rate of Couette flows.

\begin{figure}
\centering
\includegraphics{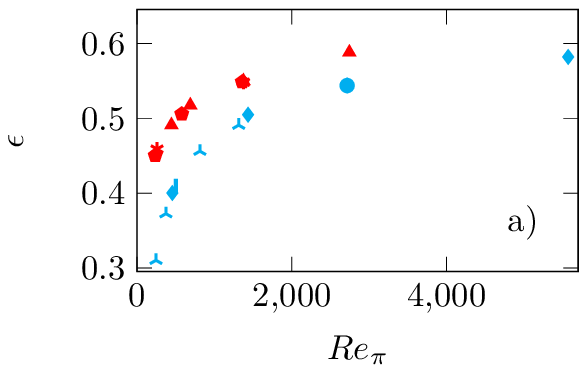}
\includegraphics{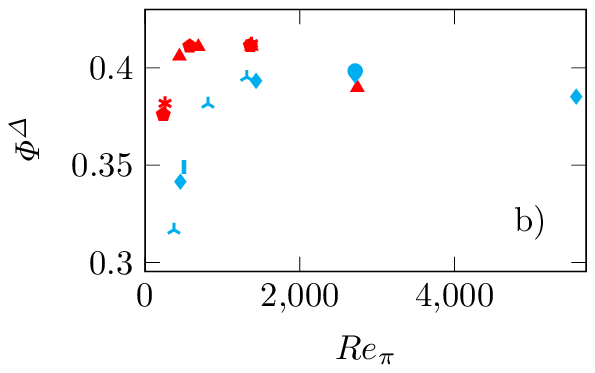}
\caption{Turbulent dissipation $\epsilon$ and deviational dissipation $\phid$ versus $\repi$. For colors and symbols, refer to table \ref{tab::DNS_inventory}.}
\label{fig::constant_power-fluxes}
\end{figure}

The remaining flux terms of the TKE energy box are plotted in figure \ref{fig::constant_power-fluxes}. Here the data points follow qualitatively similar curves for the two flows, but with significant quantitative differences. The TKE dissipation $\epsilon$ is found to monotonically increase in both flows, in agreement for example with \cite{abe-antonia-2016}, even though Couette flows yield significantly larger values than Poiseuille ones. As for the deviational dissipation $\phid$, \cite{gatti-etal-2018} observed an increasing trend for Poiseuille flows until a maximum at intermediate $Re_\pi$ is reached; then, the curve decreases monotonically to asymptotically approach zero. The same trend is here confirmed for Couette flows, except that the maximum is reached much earlier than in Poiseuille ones, supporting once more the notion that the former develop faster with $Re$. Moreover, the available Couette data points are consistently above those for Poiseuille flows, except the one at the highest $Re$; a trend reversal is thus possible at high Reynolds number, even though additional high-$Re$ Couette data would be needed to establish it properly.

\subsection{Comparison at constant $\realp$}
\label{sec::const_realp}


Additional considerations can be made when the two flows are compared at the same value of $\realp$. It has already been shown how this is equivalent to enforcing an identical MKE box, and that $\realp$ significantly differs for the two flows at the same $h^+$ or $Re_\pi$. 


\begin{figure}
\centering
\includegraphics{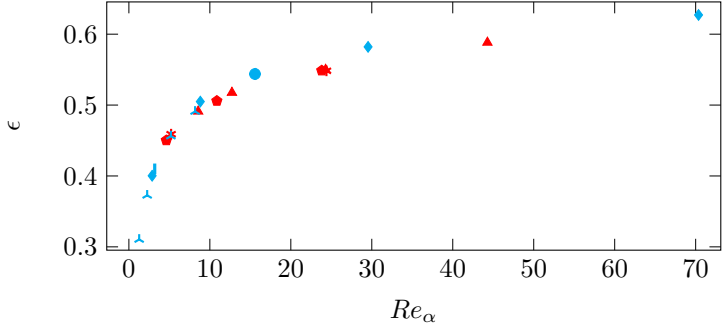}
\caption{Turbulent dissipation $\epsilon$ against $Re_\alpha$. For colors and symbols, refer to table \ref{tab::DNS_inventory}.}
  \label{fig::diss_realp}
\end{figure}

Figure \ref{fig::diss_realp} plots the turbulent dissipation $\epsilon$ against $\realp$. A single point on this plot, i.e. a pair of $(\realp, \epsilon)$ values, is sufficient to determine the whole energy box. Indeed, the value of $\realp$ sets both $\ppl$ and $\phil$, while the value of $\epsilon$ provides the missing information to recover deviational dissipation from equation \eqref{eq::tke_bal} as well as deviational production, since $\phid = -\ppd$.

The two curves for Poiseuille and Couette flows in figure \ref{fig::diss_realp} are very similar, not only qualitatively but also quantitatively. This is quite remarkable, since in principle nothing prescribes such different flows with different mechanisms of power input and at different $h^+$ and $Re_\pi$ to redistribute the same power fraction $\ppl$ identically between $\epsilon$ and $\phid$. A striking difference with respect to figure \ref{fig::constant_power-fluxes}(a) is that, here, Couette flows show a lower turbulent dissipation than Poiseuille ones, albeit marginally. Moreover, the present data indicates a tendency for this difference to increase with $Re$. In other words, when the same fraction of power $\ppl$ is transferred to the field of turbulent fluctuations, a Couette flow dissipates less of it as turbulent dissipation (and consequently more of it as $\phid$), in a manner that becomes more evident for increasing $Re$. This trend occurs despite both flows achieve $\epsilon \rightarrow 1$ (thus $\phid \rightarrow 0$) in the limit $\realp \rightarrow \infty$, which is then approached at a faster rate in Poiseuille flows. This can be readily shown through equation \eqref{eq::expr_phid} under the asymptotic-$Re$ assumption that $r(y)$ equals the total shear stress $\tau_w\psi(y)$.

\subsection{Discriminating large- and small-scale dissipation}

The above observation can be explained by the role of the large-scale structures in the two flows; as already stated, such structures are more intense in Couette than in Poiseuille flows, at least for the relatively wide range of Reynolds numbers observed in literature. The lower turbulent dissipation in Couette flows might thus be attributed to the reduced ability of these large-scale motions to transform $\ppl$ into $\epsilon$. Moreover, the implied larger $\phid$ is a sign of the large-scale motions being more efficient at producing Reynolds shear stresses, consistently with observations by \cite{lee-moser-2018}. Since the shear stress is directly related to the mean flow by the mean momentum equation, large scales affect the mean flow more than small ones.

To confirm this hypothesis, a decomposition of the fluctuating velocity field into large- and small-scale components is carried out. The procedure closely follows the one devised by \cite{kawata-alfredsson-2018}. A large-scale field $\mathbf{u}^\ell$ is defined via a sharp Fourier filter in the homogeneous directions, and the small-scale field is consequently defined as $\mathbf{u}^s = \mathbf{u}' - \mathbf{u}^\ell$. Budget equations were derived by \cite{kawata-alfredsson-2018} for the kinetic energy of these two fields. These equations resemble the one for the kinetic energy of the whole fluctuation field, and feature equivalent terms, plus a key additional transport term that describes the energy transfer between the large and small scales. The cross-scale transport term is conventionally defined to be positive when the large-scale field is receiving power. The equation for the large-scale kinetic energy becomes:
\begin{eqnarray}
\frac{\bar{D}}{Dt} \frac{ \aver{u_i^\ell u_i^\ell}}{2} =  
\underbrace{\shortminus \aver{ u^\ell v^\ell }\frac{dU}{dy}}_{\text{production}} +
\underbrace{\shortminus \frac{1}{2}\frac{d}{d y} \aver{v' u_i^\ell u_i^\ell}}_{\text{turb. transport}} + 
\underbrace{\shortminus \aver{u_i^\ell u'_k \frac{\partial u_i^s}{\partial x_k}}}_{\text{cross-scale transport}} + 
\underbrace{\shortminus\frac{d}{dy}\aver{P' v^\ell}}_{\text{press. transport}} \notag \\ + \underbrace{\frac{1}{\re}\frac{d^2}{dy^2}\frac{\aver{u_i^\ell u_i^\ell}}{2}}_{\text{viscous diffusion}}	- 
\underbrace{\frac{1}{\re}\aver{\frac{\partial u_i^\ell}{\partial x_k} \frac{\partial u_i^\ell}{\partial x_k}}}_{\text{turb. dissipation}}
\label{eq::large_scale_local}
\end{eqnarray}
and substituting $\mathbf{u}^\ell$ with $\mathbf{u}^s$ provides the analogous equation for the small-scale field. The crucial cross-scale term describes the energy exchange between small and large scales, and represents a non-local process in the physical space, so that the exchange only balances after integration on the whole domain, i.e.
\begin{equation}
\int_{-1}^1  
\underbrace{\shortminus\aver{u_i^\ell u'_k \frac{\partial u_i^s}{\partial x_k}}}_{\text{cross-scale, large}} +
\underbrace{\shortminus\aver{u_i^s u'_k \frac{\partial u_i^\ell}{\partial x_k}}}_{\text{cross-scale, small}} dy = 0
\end{equation}

Volume-integration of the budget equation \eqref{eq::large_scale_local} and its small-scale equivalent yields the energy fluxes $\ppl$, $\ppd$ and $\epsilon$ separated into their small- and large-scale components, which will be indicated by subscripts $\cdot_s$ and $\cdot_\ell$ respectively, plus the interscale transfer $T$.


We compute the decomposed energy budget for two Couette flows at $h^+\simeq 100$ and $500$. As in \cite{kawata-alfredsson-2018}, no filtering is carried out in the streamwise direction, owing to the very elongated nature of the large-scale structures. For spanwise filtering, the selection of the wavelength $\lambda_z$ to discriminate the large-scale motion is guided by the rather flat peak of the energy spectrum observed in \cite{lee-moser-2018} in the range $3 < \lambda_z < 6.5$. Detailed scrutiny of the same spectrum for the present datasets has determined the range of interest to be $\lambda_z \geq \pi$, which is used as a criterion to discriminate the large scales from the small ones.

\begin{figure}
\centering
\includegraphics[]{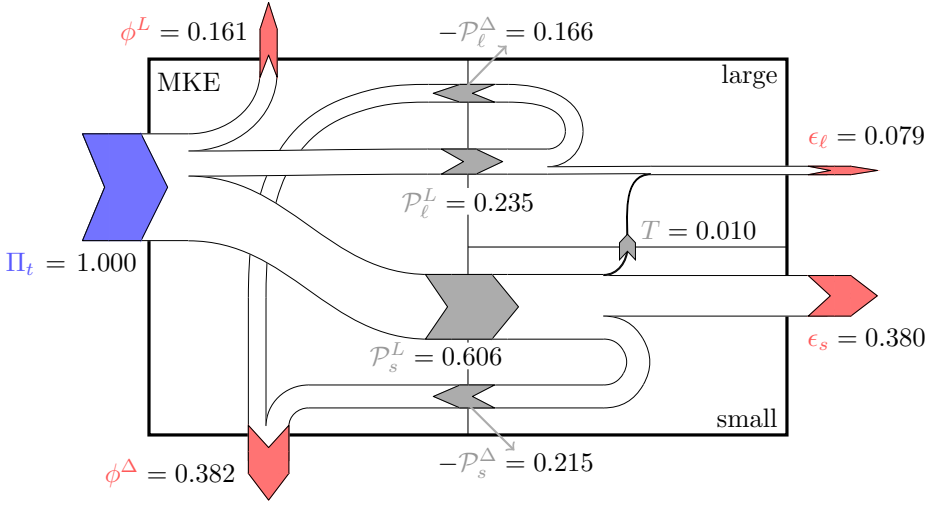} 
\caption{Extended energy box for a Couette flow at $h^+=102$, with the TKE box split into large- and small-scale contributions.}
\label{fig::tennis}
\end{figure}

\begin{figure}
  \centering
  \includegraphics[]{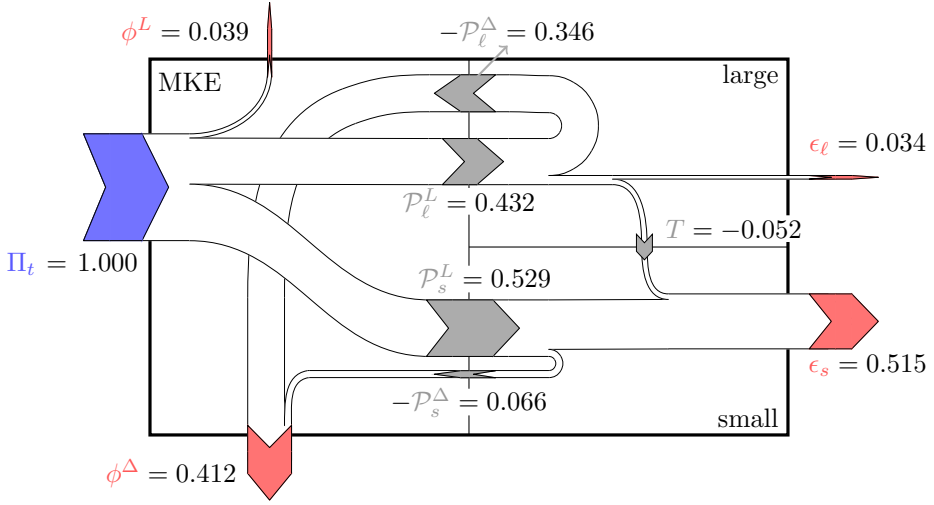} 
  \caption{Extended energy box for a Couette flow at $h^+=507$, with the TKE box split into large- and small-scale contributions.}
  \label{fig::tennis_500}
\end{figure}

The decomposed energy budget is given in figure \ref{fig::tennis} for the Couette flow at $h^+\simeq100$, where the TKE box is separated into two sub-boxes pertaining to the small and large scales. It is confirmed that the large scales are less efficient than the small ones at producing $\epsilon$, with $\epsilon_\ell=0.079$ versus $\epsilon_s=0.380$. Not only do small scales produce more turbulent dissipation, but they also convert a larger fraction of the received energy into it, as $\epsilon_\ell/\ppl_\ell = 0.336$ while $\epsilon_s/\ppl_s = 0.627$. Obviously, their impact on the deviational dissipation is opposite, with $|\ppd_\ell| / \ppl_\ell=0.706$ and $\ppd_s/\ppl_s=0.355$. The same picture is observed in figure \ref{fig::tennis_500} for a Couette flow at $h^+ \simeq 500$, with even more pronounced features because of the higher $Re$ and the increased separation between large and small scales. The large-scale contribution to turbulent dissipation becomes much smaller than the small-scale one ($\epsilon_\ell = 0.034$ against $\epsilon_s=0.515$); moreover, the large scales account for the most of the deviation dissipation with $|\ppd_\ell|/\phid = 0.8398$. In other words, turbulent dissipation is clearly dominated by small-scale effects, whereas deviation dissipation is mainly caused by large-scale ones.



In both cases, the interscale net energy flux is rather small: $T=0.01$ for $h^+ \simeq 100$ and $T=-0.052$ for $h^+ \simeq 500$. However small, the net integral effect at the lower Reynolds number is to transfer energy from the small to the large scales. The same result was obtained by \cite{papavassiliou-hanratty-1997} by treating large-scale structures as secondary motions; this differs however from the more recent observation put forward by \cite{kawata-alfredsson-2018}, where an inverse interscale transport was only found for the Reynolds shear stress $r(y)$. At higher Reynolds number the direction of the transfer is reversed, and power goes from large to small scales. This inversion can be explained by figure \ref{fig::profile_tcross}, that plots the $y$-profile of the cross-scale transport term of the large scales, i.e.
\begin{equation}
  \shortminus\aver{u_i^\ell u'_k \frac{\partial u_i^s}{\partial x_k}}\,\text{.}\notag
\end{equation}
A positive peak is present in the near-wall region, where the large scales are receiving power; the present data suggest that its position scales in viscous units. Hence the peak covers a large portion of the domain at low Reynolds number -- meaning that the integral flux $T$ is dominated by near-wall effects, and ends up being positive at $h^+ \simeq 100$. Conversely, at higher $h^+$ the integral is dominated by the core region of the flow, where the large scales lose energy, thus explaining the negative value of $T$ at $h^+ \simeq 500$. This interpretation agrees e.g. with findings by \cite{cho-hwang-choi-2018} and \cite{kawata-tsukahara-2021}, who detected an inverse energy cascade from large energy-containing motions to even larger ones in the proximity of the wall.

\afterpage{
  \begin{figure}
    \centering
    \includegraphics{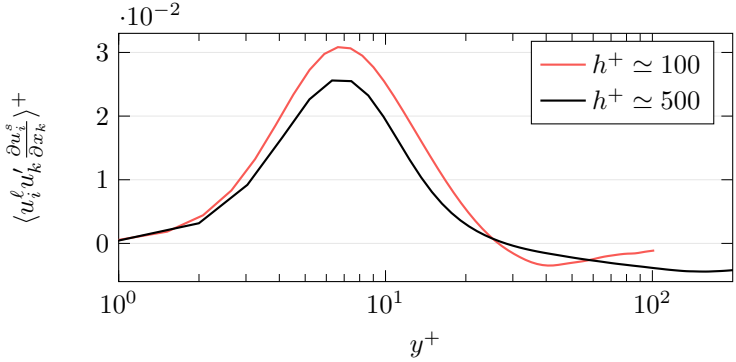}
    \caption{Profiles of the cross-scale transport term for the large-scale fluctuation field $\mathbf{u}^\ell$ in Couette flows, plotted in viscous units. Red: $h^+ = 102$; black: $h^+ = 507$.}
    \label{fig::profile_tcross}
  \end{figure}
}

\section{Conclusions}
\label{sec::conclusions}

Turbulent plane Poiseuille and Couette flows have been compared in terms of their integral energy fluxes, thanks to the analysis of a database of direct numerical simulations. The work is based on the framework introduced by \cite{gatti-etal-2018} to compare two Poiseuille flows with and without flow control under the same power input (i.e. at Constant Power Input, or CPI). The CPI approach is extended here to the case of a generic plane parallel flow driven by both shear and a pressure gradient. A power-based velocity scale $U_\pi^*$ and a corresponding Reynolds number $\repi$ are defined and used for non-dimensionalisation, so that two flows with the same power input possess the same $U_\pi^*$ and the same $\repi$. After a standard Reynolds decomposition, the mean flow is further split into a laminar (Stokes)  and a mean deviation contribution. This procedure also decouples all the volume-integrated energy fluxes into a laminar and a deviational part. The extended decomposition, together with normalization by the total power input, enables expressing all the energy fluxes as functions of the sole variables $\repi$ and two wall-normal integrals of the Reynolds shear stress.

Comparing Couette and Poiseuille flows at the same $\repi$ ascertains that Couette ones produce a slightly larger flow rate, i.e. Couette flows are more effective at converting a given input power into flow rate. However, the efficiency of the process is a completely different concept. Among the volume-integrated fluxes, the flux $\ppl$ (laminar production of turbulent kinetic energy) indicates the total fraction of power that is wasted as an overhead expense owing to the presence of turbulence, or the flow (in)efficiency. The flux $\phil = 1 - \ppl$ is the laminar dissipation, i.e. the dissipation of the laminar flow, and similarly expresses the efficiency of the flow, being the ratio between the theoretical minimum power necessary to realize a given flow rate and the actual power used to drive the turbulent flow. For a comparison under the CPI condition, Couette flows produce a larger $\ppl$ (see figure 4) and a lower efficiency $\phil=1-\ppl$. The two observations that Couette flows are at the same time more effective (i.e. produce a larger flow rate for a given power) and less efficient (i.e. waste a larger power share to turbulence) are only apparently contradictory: the laminar Couette solution requires less power than its Poiseuille counterpart to produce the same flow rate, hence compensating for the higher turbulent activity of the former. The Stokes solution is therefore quite relevant in determining the absolute performance of a turbulent flow at producing flow rate. Also, the efficiency $\phil$ should not be used to compare effectiveness at producing flow rate across different flows, but only to indicate how the flow compares to the ideal situation.

In the case of simple Couette and Poiseuille flows, both fluxes $\phil$ and $\ppl$ have been written as functions of a sole variable: the Reynolds number $\realp$, which is embedded in the FIK identity. Its velocity scale is the ratio between a weighted integral of the shear stress and the flow rate. Surprisingly, these two functions of $\realp$ turn out to be identical for Couette and Poiseuille flows. However, the same value of $\realp$ is achieved by Couette flows at a lower value of $h^+$, explaining and -- more importantly -- quantifying how turbulence in Couette flows develops faster with the Reynolds number.

Couette and Poiseuille flows can also be compared at the same value of $\realp$, which corresponds to a situation where the fluctuating field is fed with the same fraction $\ppl$ of the total power. In this case, Couette flows are found to achieve a smaller turbulent dissipation $\epsilon$, even though for both flows $\epsilon$ tends to unity at infinite Reynolds number. Indeed, Couette flows feature stronger large-scale structures, which are efficient at modifying the mean flow, but carry a lower contribution to turbulent dissipation compared to smaller scales. To verify this, an inter-scale analysis of the energy fluxes has been performed on two Couette flows at $h^+\simeq100$ and $500$. The TKE integral budget is further divided in small- and large-scale contributions, with an additional transport term that quantifies the cross-talk between scales, as shown in figure \ref{fig::tennis} and \ref{fig::tennis_500}. By setting the scale separation threshold in such a way that the large rolls are included into the large-scale part, it is found that the large scales produce only a minor fraction of the total $\epsilon$ (approximately $10-20\%$ of it); moreover, they convert to turbulent dissipation only a small fraction (approximately $10-30\%$) of the power $\ppl_\ell$ they are fed with. This behaviour is opposite to the one of the small-scales, and becomes more pronounced as the Reynolds number increases. At low Reynolds numbers, the interscale transport term is found to move energy from the small to the large scales; the opposite happens at higher Reynolds number. Such an inverse interscale transport was indeed already observed by \cite{kawata-alfredsson-2018} but limited to the shear stress, whereas other studies \citep{kawata-tsukahara-2021,cho-hwang-choi-2018} have already reported a reversed energy cascade in proximity of the wall. Our data suggest that this near-wall, inverse-cascade region scales in viscous units; hence it contributes more to the cross-talk at lower Reynolds numbers, and becomes less important as Reynolds increases.

The framework proposed here for the analysis of global energy budgets in turbulent flows, by also accounting for the presence of large-scale structures, has been demonstrated for the Couette--Poiseuille family of flows, but its usability is larger. It can be applied to unconventional cases such as Poiseuille flows with zero flow rate \citep{tuckerman-etal-2014} or Couette flows in a rotating reference frame \citep{kawata-alfredsson-2019,bech-andersson-1996}. Every term of the energy box except $\epsilon$ (and $T$ in the scale-decomposed version) can be obtained from the profile of the Reynolds shear stress, hence the framework suits experimental studies as well. Flow control schemes can also be assessed, as done in \cite{gatti-etal-2018}, providing useful insights on the role played by large-scale motions in the context of drag reduction. \cite{roccon-zonta-soldati-2021} have recently and successfully applied it to two-phases flows with complex physics. Flows featuring large or secondary motions, such as open channel flows \citep{zampiron-etal-2021}, are of interest as well; the same applies to to straight duct flows with arbitrary geometry. Finally, as pointed out by \cite{frohnapfel-hasegawa-quadrio-2012}, generalisation to external flows such as spatially developing boundary layers is also possible.

\backsection[Acknowledgements]{The authors acknowledge support by the state of Baden-W\"urttemberg through bwHPC, which provided computational resources.}

\backsection[Funding]{This work is supported by the Priority Programme SPP 1881 Turbulent Superstructures of the Deutsche Forschungsgemeinschaft (A.A. and D.G., grant number GA 2533/1-1).}

\backsection[Declaration of interests]{The authors report no conflict of interest.}


\end{document}